\documentclass[preprint,authoryear,12pt]{elsarticle}

\usepackage{amssymb}
\usepackage{natbib}
\usepackage{amsmath,amsfonts,amsthm,bm}
\usepackage{graphicx}
\usepackage{lineno} 
\usepackage{float}
 \usepackage{setspace}
\usepackage[nodots]{numcompress}
\usepackage[top=1in, bottom=1in, left=1.25in, right=1.25in]{geometry}

\usepackage{algorithm}
\usepackage{algorithmic}
\usepackage{breqn}
\usepackage{booktabs,caption}
\usepackage[flushleft]{threeparttable}
\usepackage{tabularx}

\usepackage{caption}
\usepackage[skip=5pt]{caption}
\usepackage[colorlinks]{hyperref}
\usepackage[nameinlink]{cleveref}
\Crefname{figure}{Fig.}{Figs.}% {<type>}{<singular>}{<plural>}
\newcommand{\Lagr}{\mathcal{L}}

\begin{document}
\begin{spacing}{1.25}

\begin{frontmatter}

\title{\textbf{An advanced hybrid deep adversarial autoencoder for parameterized nonlinear fluid flow modelling}}

\author[label1]{\textbf{M.Cheng}}

\author[label1]{\textbf{F.Fang}\corref{cor1}}
\address[label1]{Applied Modelling and Computation Group, Department of Earth Science and Engineering, Imperial College London, SW7 2BP, UK}

\cortext[cor1]{Corresponding author}
\ead{f.fang@imperial.ac.uk}

\author[label1]{\textbf{C.C. Pain}}

\author[label2]{\textbf{I.M.Navon}}
\address[label2]{Department of Scientific Computing, Florida State University, Tallahassee, FL, 32306-4120, USA}

\begin{abstract}
Considering the high computation cost produced in conventional computation fluid dynamic simulations, machine learning methods have been introduced to flow dynamic simulations in recent years. However, most of studies focus mainly on existing fluid fields learning, the prediction of spatio-temporal nonlinear fluid flows in varying parameterized space has been neglected. In this work, we propose a hybrid deep adversarial autoencoder (DAA) to integrate generative adversarial network (GAN) and variational autoencoder (VAE) for predicting parameterized nonlinear fluid flows in spatial and temporal space. High-dimensional inputs are compressed into the low-representation representations by nonlinear functions in a convolutional encoder. In this way, the predictive fluid flows reconstructed in a convolutional decoder contain the dynamic flow physics of high nonlinearity and chaotic nature. In addition, the low-representation representations are applied into the adversarial network for model training and parameter optimization, which enables a fast computation process. The capability of the hybrid DAA is demonstrated by varying inputs on a water collapse example. Numerical results show that this hybrid DAA has successfully captured the spatio-temporal flow features with CPU speed-up of three orders of magnitude. Promising results suggests that the hybrid DAA can play a critical role in efficiently and accurately predicting complex flows in future.
\end{abstract}


\end{frontmatter}

%% main text
%\linenumbers

\section{Introduction}
\label{sec1} 
Numerical simulations of nonlinear fluid flows are used to describe the complex evolution of many physical processes. Accurate simulations of nonlinear fluid flows are of great significance to many fields such as flood and ocean modelling, which systems often exhibit rich flow dynamics in both space and time \citep{ erichson2019physics, hu2019rapid}. The numerical simulations have benefited from the availability of high-resolution spatio-temporal data with recent advances in measurement techniques \citep{ taira2019modal}, which makes the studies of more complex flows become possible. However, the computational cost involved in solving complex problems is intensive which still precludes the development in these areas. In order to address the issue of high computational cost, this paper proposes a hybrid deep adversarial autoencoder to solve fluid problems in an efficient manner.

Recent advances in machine learning technologies are increasingly of interest for efficiently simulating flows in dynamical systems \citep{lusch2018deep, lee2018model, murata2019nonlinear}. Machine learning has demonstrated its potential capability in fluid flow applications \citep{Brunton2019MachineLF}, such as fluid flow modelling \citep{geneva2019quantifying, ling2016reynolds}, flow simulation \citep{kutz2017deep, liu2017flood}, and fluid field reconstruction \citep{farimani2017deep, kim2019deep}. For example, \cite {humphrey2016hybrid} utilized a Bayesian artificial neural network (ANN) with a conceptual model to forecast monthly streamflow. \cite{mohan2018deep} developed the Long Short Term Memory (LSTM), a type of recurrent neural network (RNN), to simulate the temporal dynamics of turbulent flows. \cite{kim2019deep} adopted a convolutional neural network (CNN) to reconstruct fluid fields. Despite these researches demonstrated to successfully reconstruct flow dynamics, it is noted that they commonly do not take into account the temporal and spatial evolution of inputs or parameters, which is crucial for realistic dynamical systems \citep{reichstein2019deep}. 

Most recently, the generative adversarial networks (GANs) have been developed for predicting the parameterised nonlinear fluid flows \citep{farimani2017deep}. GAN introduced by \cite{goodfellow2014generative}, recently has emerged as a leading role for recreating the distributions of complex data \citep{xie2018tempogan}. The key feature of GAN is the adversarial strategy in two modules. GAN defines a learning generative network by transforming a latent variable into a state variable using nonlinear functions. Then GAN drives the learning process by discriminating the observed data from the generated data in a discriminator network. Because of the special adversarial architecture, GAN has demonstrated great capability in producing high-resolution samples, mostly in images, e.g., image synthesis \citep{reed2016generative}, semantic image editing \citep{li2016precomputed}, style transfer \citep{isola2017image} etc.

For efficient GAN training and accurate spatio-temporal fluid flow prediction, Variaitonal Autoencoder (VAE) \citep{kingma2013auto} has been introduced to GAN in this study. VAE has been widely used in various research areas, such as text Generation \citep{ semeniuta2017hybrid}, facial attribute prediction \citep{hou2017deep}, image generation \citep{ walker2016uncertain, pu2016variational}, graph generation \citep{simonovsky2018graphvae}, music synthesis \citep{roberts2017hierarchical}, and speech emotion classification \citep{ latif2017variational}, etc. The hybrid deep adversarial autoencoder (DAA) developed here takes advantages of both GAN and VAE. GAN allows
for training on large datasets and is fast to yield visually and high-resolution images, but the flexible architecture is easy to come with the model collapse problem and generate unreal results \citep{rosca2017variational}. The VAE is attractive for achieving better log-likelihoods than GAN \citep{wu2016quantitative, mescheder2017adversarial}, therefore it encourages the hybrid DAA to better represent all the training data and discouraging mode-collapse problem in GAN \citep{rosca2017variational}. 

The hybrid DAA developed here is a robust and efficient numerical tool for accurate prediction of parameterised nonlinear fluid flows. The advantages of the hybrid DAA include:
\begin{itemize}
\item The proposed method exploits spatial features by use of convolutional neural networks. It will be advantageous over the traditional reduced order models (ROMs) \citep{fang2017efficient, xiao2019domain} since the high-dimensional datasets are compressed into the low-dimensional representations by nonlinearity functions in a convolutional encoder. In this way, the predictive fluid flows containing high nonlinearity and chaotic nature can be represented by a convolutional decoder.
\item The low-dimensional representations, several orders of magnitude smaller than the dimensional size of the original datasets, are applied into the adversarial network for representation learning and parameter optimization, thus accelerating the computation process. 
\item With the trained hybrid DAA, for any given different inputs, the spatio-temporal features can be automatically extracted in the encoder. Consequently, accurate predictive nonlinear fluid fields can be further obtained in the decoder with an efficient manner.
\end{itemize}
This is the first time that the hybrid DAA is adopted to address parameterised nonlinear fluid flow problems. It will make a breakthrough in predicting accurate nonlinear fluid flows with the high-speed computation.

The reminder of this paper is organised as follows. Methodologies of VAE and GAN are briefly introduced in \cref{sec2} and \cref{sec3}, respectively. The hybrid DAA for parameterised nonlinear fluid fields is detailed described in \cref{sec4}. \cref{sec5} demonstrates the performance of the hybrid DAA using water collapse as a test case. Finally in \cref{sec6}, conclusions are presented.

\section{Variational Autoencoder }
\label{sec2}
Variational Autoencoder (VAE) is introduced by \citet{kingma2013auto}, which combines Bayesian inference with deep learning. The VAE is a generative model which aims to produce the desired local variable ${\hbar}$ from the underlying latent variable ${\zeta}$. Mathematically, let ${p_{\zeta}}(\zeta )$ be a prior distribution of ${\zeta}$, and the probability of the local state variable ${\hbar}$ be modelled by
\begin{equation}
\label{eq1}
{\hbar \sim {p _\theta }(\hbar |\zeta )},{p}(\hbar ) = \int {p_\theta }(\hbar |\zeta ){p_{\zeta}}(\zeta ) {d\zeta },
\end {equation}
where ${{p_\theta }(\hbar |\zeta )}$ is the conditional distribution of the local state variable ${\hbar}$ given ${\zeta}$, which is modelled by deep neural networks (called decoder) with parameters ${\theta}$. 

In contrast to standard autoencoders, the key property of VAE is the ability to control the distribution of the latent state vector $\zeta$, which is usually modelled by a standard Gaussian distribution ${\mathcal{N} (\zeta | 0; I) }$ \citep{kingma2013auto}. In VAE, the probability of the latent vector $\zeta$ can be expressed as
\begin{equation}
\label{eq2}
{\zeta \sim {q_\phi }(\zeta |\hbar)}, 
\end {equation}
where ${{q_\phi }(\zeta |\hbar)}$ is the conditional distribution of the local state variable ${\hbar}$ given ${\zeta}$, which is modelled by deep neural networks (called encoder) with parameters ${\phi}$.

To achieve the sample reconstruction, the reconstruction loss ${{\Lagr _{rec}} }$ as the negative expected log-likelihood of the samples needs to be maximized, as following:
\begin{equation}
\label{eq3}
{\Lagr _{rec}} = {E_{{q_\phi }(\zeta |\hbar )}}(\log {p_\theta }(\hbar |\zeta )).
\end{equation}

The difference (called Kullback–Leibler divergence) between the distribution of ${q(\zeta |\hbar )}$ and a prior distribution (for example, Gaussian distribution) ${p_{\zeta}}(\zeta )$ = ${\mathbb{N} (\zeta | 0; I) }$ can be quantified as
\begin{equation}
\label{eq4}
{\Lagr _{KL}} = D_{KL}({q_\phi }(\zeta |\hbar )||{p_\theta}(\zeta)).
\end{equation}

The total loss is consisted of the reconstruction loss and the Kullback–Leibler (KL) divergence (${\Lagr _{vae} = {\Lagr _{rec}}+ {\Lagr _{KL}}}$), which can be minimized by gradient descent algorithms \citep{kingma2013auto}. Since the KL divergence is non-negative, the total loss can be expressed as:
\begin{equation}
\label{eq5}
{\Lagr _{vae}} = - D_{KL}({q_\phi }(\zeta |\hbar )||{p_\theta }(\zeta )) + {E_{{q_\phi }(\zeta |\hbar )}}(\log {p_\theta }(\hbar |\zeta )).
\end{equation}

Correspondingly, the full objective function of VAE is as follows
\begin{equation}
\label{eq6}
\mathop {\max }\limits_\theta \mathop {\max }\limits_\phi {E_{{p_{data}}(\hbar)}}[ - D_{KL}({q_\phi }(\zeta |\hbar )||{p_\theta }(\zeta )) + {E_{{q_\phi }(\zeta |\hbar )}}(\log {p_\theta }(\hbar |\zeta ))],
\end{equation}
where ${p_{data}(\hbar)}$ is the prior distribution of ${\hbar}$.

\section{Generative adversarial network}
\label{sec3}
The GAN is generally implemented with a minimax game in a system of two players \citep{goodfellow2016nips}. One player is responsible for generating the new samples ${\hbar}$ from a random dataset ${\mu}$ , while another player aims to discriminate the real samples ${\hbar_d}$ from the generated samples ${\hbar}$. The former player is called the generator ${\mathcal G}$ and the latter is the discriminator ${\mathcal D}$.

In the discriminator, ${\mathcal D (\hbar_d)=1}$ if the real samples ${\hbar_d}$ are accepted while ${\mathcal D (\mathcal G (\mu)) =0}$ if the generated samples ${\hbar}$ (${\hbar = \mathcal G (\mu)}$) rejected. Unlike the autoencoder, GAN is a two-player game rather than optimizing one loss function ${\Lagr _{vae}}$ (as in Eq.\eqref{eq6}) in VAE. During the training process of GAN, the parameters in the discriminator are updated by maximizing ${{\Lagr_D}}$ as:
\begin{equation}
{\Lagr_D} = {E_{{\hbar_d}\sim{p_{data}(\hbar_d)}}}[\log {\mathcal D}({\hbar _d})] + {E_{\mu \sim{p_{\mu}(\mu)}}}[\log (1 - {\mathcal D}({\mathcal G}(\mu )))],
\end{equation}
while the parameters in the generator are updated by minimizing ${{\Lagr_G}}$ as:
\begin{equation}
{\Lagr_G} = {E_{\mu  \sim {p_{\mu}(\mu)}}}[\log (1 - {\mathcal D}({\mathcal G}(\mu )))],
\end{equation}
where ${p_{\mu}(\mu)}$ is a prior distribution for the random dataset ${\mu}$, and ${p_{data}({\hbar_d})}$ is the corresponding probability data distribution for the real datasets ${\hbar_d}$. 

Concretely, the objective function for GAN is shown in Eq.\eqref{eq:3}:
\begin{equation}
\label{eq:3}
\mathop {\min }\limits_{\mathcal G} \mathop {\max }\limits_{\mathcal D} {\Lagr }({\mathcal G},{\mathcal D}) = {E_{{\hbar_d}\sim{p_{data}(\hbar_d)}}}[\log {\mathcal D}({\hbar_d})] + {E_{\mu \sim{p_{\mu}(\mu)}}}[\log (1 - {\mathcal D}({\mathcal G}(\mu )))],
\end{equation}

In practice, this optimization of ${{\Lagr }({\mathcal G},{\mathcal D})}$ is performed alternately using gradient-based methods, e.g. the adaptive Moment Estimation (Adam) optimizer \citep{kingma2014adam}. Given enough capacity, the game converges to a global optimum where ${{\mathcal D }(\hbar_d) = \frac{1}{2}}$ everywhere (details shown in \citet{goodfellow2014generative}).

\section{Hybrid deep adversarial autoencoder for nonlinear fluid flow modelling}
\label{sec4}
\begin{figure}[H]
\centering
\includegraphics[width=1.0\textwidth]{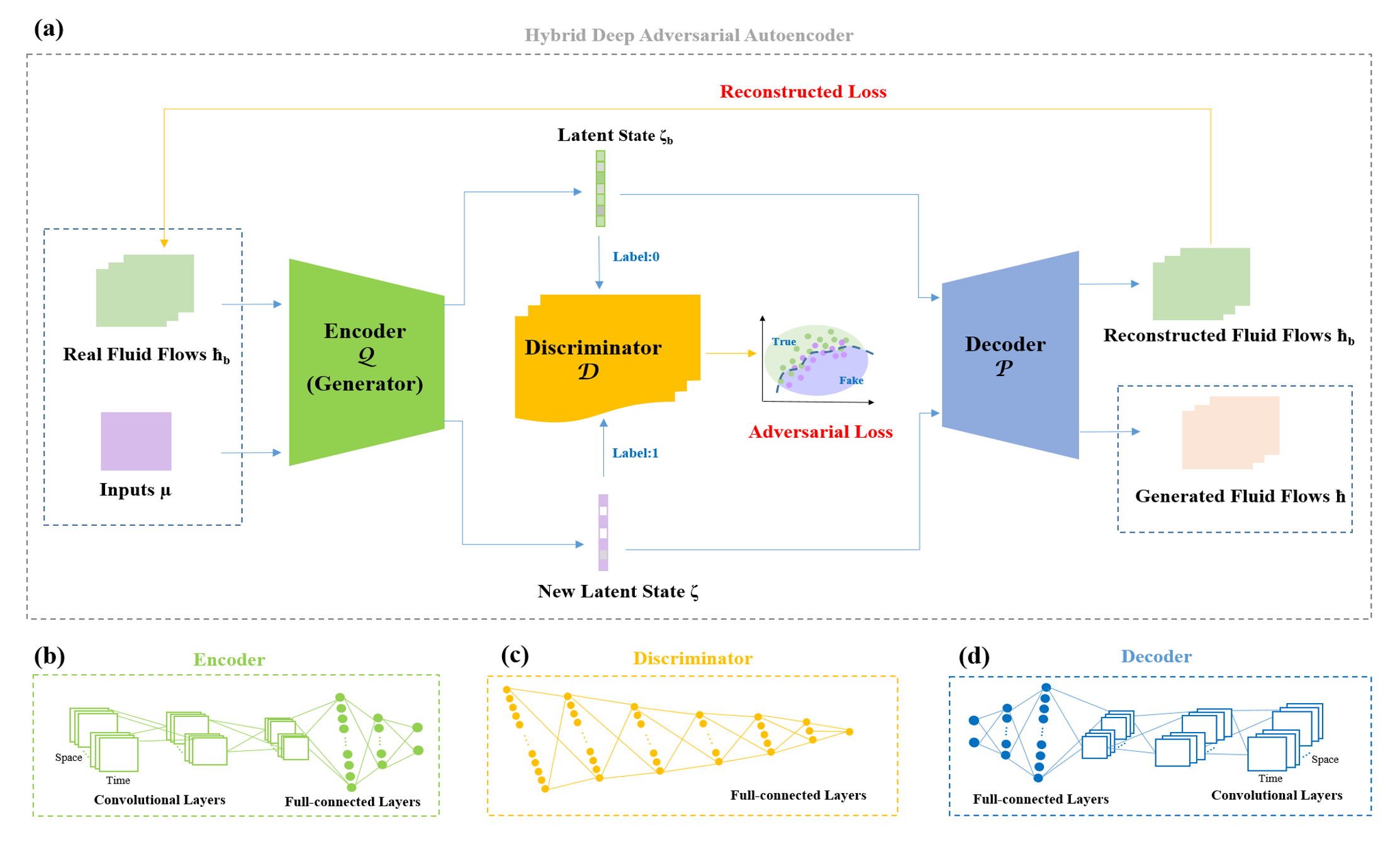}
\caption{(a) Illustration of a hybrid deep adversarial autoencoder (DAA), which consists of an encoder (acted as a generator) (b), a discriminator (c), and a decoder (d). The encoder and decoder compose of convolutional layers and full-connected layers. The discriminator is a stack of full-connected layers.}
\label{AAE}
\end{figure}

In this paper, for nonlinear fluid flow modeling, a hybrid deep learning fluid model based on deep adversarial autoencoder (DAA) is proposed which is established by combining a VAE and a GAN. As shown in \Cref{AAE}(a), the encoder is acted as a generator in the hybrid DAA. In comparison to Eq.\eqref{eq2} in VAE, the generator here defines an aggregated posterior distribution of ${q(\zeta)}$ on the latent state as followings:
\begin{equation}
\label{eq10}
{q}(\zeta ) = \int {q_\phi }( \zeta |\hbar){q_{data}}(\hbar ) {d\hbar },
\end {equation}
where ${q_{data}({\hbar})}$ is the data distribution for the datasets ${\hbar}$. The autoencoder in the hybrid DAA is regularized by matching the aggregated posterior ${q(\zeta)}$ to the prior ${p(\zeta )}$ in Eq.\eqref{eq1}.

As described in \Cref{AAE}(a), in the forward propagation process, the inputs ${\mu}$ and the targeted outputs ${\hbar_b}$ are fed into the encoder which generates the corresponding latent states ${\zeta}$ and ${\zeta_b}$ respectively. The encoder tries to produce the aggregated posterior distribution ${q(\zeta_b)}$ matched with the prior distribution ${p(\zeta)}$, which can fool the discriminator in GAN. The latent states ${\zeta_b}$ and ${\zeta}$ are then transformed into the reconstructed targeted outputs ${\hbar_b}$ and generated outputs ${\hbar}$ respectively in the decoder. In the backward propagation process, the parameters in modules: encoder, decoder and discriminator are updated by minimizing the reconstruction loss and maximizing the adversarial loss.

In general, a parameterized partial differential equation for a spatio-temporal fluid flow problem can be written as
\begin {equation}
\label{f1}
{\mathcal M} (\hbar (x,\mu ,t),x,\mu ,t) = {\mathcal F} (\hbar (x,\mu ,t),x,\mu ,t),
\end {equation}
where ${\mathcal M}$ denotes a nonlinear partial differential operator, ${\hbar (x,\mu ,t)}$ is the state variable vector (for example, pressure, density, velocity, etc), ${x}$ represents the spatial coordinate system, ${\mu}$ denotes the parameter vector (for example, model input and boundary condition), ${t}$ is the time and ${\mathcal F}$ is the source term.

In spatio-temporal fluid flow simulations, the state variable vector ${\hbar}$ represents the fluid flow distribution during a specified simulation period ${[0, t_{N_t}]}$ as
\begin{equation}
\label{f22}
\hbar  = ({\hbar _0},...,{\hbar _{{t_{{n_t}}}}},...,{\hbar _{{t_{{N_t}}}}}),
\end{equation}
where ${\hbar _{t_{n_t}}}$ is the state variable vector at time level ${ t_{n_t}}$ (${{n_t} \in [0,{N_t}]}$, ${{t_{{n_t}}} \in [0,{t_{{N_t}}}]}$, ${N_t}$ is the number of timesteps). 

In a spatio-temporal discretisation form, the state variable vector ${\hbar}$ in Eq.\eqref{f22} can be rewritten as: 
\begin{equation}
\hbar = \left[ {\begin{array}{*{20}{c}}
{\hbar _{{t_1}}^1}&{\hbar _{{t_1}}^2}& \cdots &{\hbar _{{t_1}}^{{N_x}}}\\
{\hbar _{{t_2}}^1}&{\hbar _{{t_2}}^2}& \cdots &{\hbar _{{t_2}}^{{N_x}}}\\
\vdots & \vdots &{}& \vdots \\
{\hbar _{{t_{{N_t}}}}^1}&{\hbar _{{t_{{N_t}}}}^2}& \cdots &{\hbar _{{t_{{N_t}}}}^{{N_x}}}
\end{array}} \right] = [{\hbar ^1},{\hbar ^2},...,{\hbar ^{{N_x}}}],
\end{equation}
where $N_x$ denotes the number of points in scalar grids of the computational domain $\Omega$.

The parameter vector ${\mu}$ in Eq.\eqref{f1} at a spatial space $\Omega$ can be expressed:
\begin{equation}
\mu = [{\mu ^1},{\mu ^2},...,{\mu ^{{N_x}}}].
\end{equation}

Given the parameter vector ${\mu  = [{\mu ^1},{\mu ^2},...,{\mu ^{{N_x}}}]}$ as inputs and the state variable vector ${\hbar_b = [{\hbar_b ^1},{\hbar_b ^2},...,{\hbar_b ^{{N_x}}}]}$ as the targeted outputs,  the model architectur of the hybrid DAA are shown in \Cref{AAE}.

\section{Numerical examples}
\label{sec5}
The example used for validation of the hybrid DAA is a case of water column collapse. It is a benchmark test case for multi-material models, which is also known as a dam break problem. The dam-break flow problem has been of great importance in hydraulic engineering and hydropower generation. Dam-break flows are complex and generally discontinuous with abrupt variations of flow velocity and water depth \citep{seyedashraf2017dam}. The flooding induced by the dam-break flows causes great loss of human life and property as well as damaging the ecosystem in the downstream area. 

\begin{figure}[H]
\centering
\includegraphics[width=0.8\textwidth]{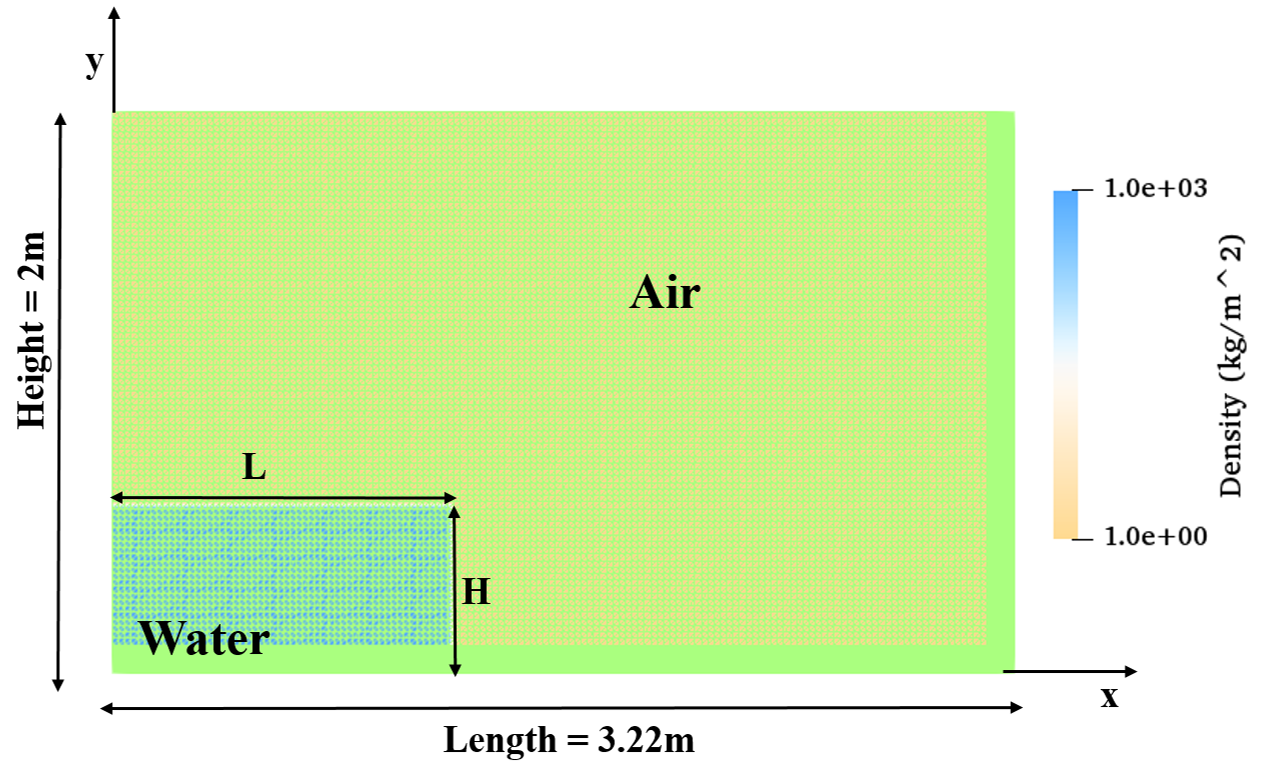}
\caption{The sketch of water collapse experiment.}
\label{Mesh}
\end{figure}

In this example, the dam-break experiment conducts in a tank with the length ${3.22 m}$, the height ${2 m}$, and the depth ${1 m}$ \citep{zhou1999nonlinear}. The simulated reservoir of water is held behind a barrier at one end of the tank. For no variations are introduced in the third dimension, the experiment is reproduced in the horizontal and vertical dimensions within the domain area ${\Omega}$. \Cref{Mesh} shows the mesh in the domain area ${\Omega}$. The dam break problem is simulated using the unstructured mesh finite element fluid model (${Fluidity}$) \citep{pain2001tetrahedral}(referred as the original high fidelity model). The densities of water and air are ${1,000kg{m^{ - 2}}}$ and ${1kg{m^{ - 2}}}$ respectively. The scalar fields ${a_k}$ representing the volume fraction is introduced to distinguish the two materials. As shown in \Cref{Mesh}, the interface between the water (the yellow area, ${a_k=1}$) and air (the blue area, ${a_k=0}$) is delineated by contours at ${a_k}$ of $0.025$, $0.5$ and $0.975$. In this case, the simulation period is ${[0, 1.85] s}$, with a timestep = ${0.025s}$. Thus, the targeted fluid flows ${\hbar_d}$ were obtained by running the original high fidelity model, with a unstructured mesh of 19097 nodes.

\subsection{Model appplication}
\label{sec5.1}

\begin{figure}[H]
\centering
\includegraphics[width=0.8\textwidth]{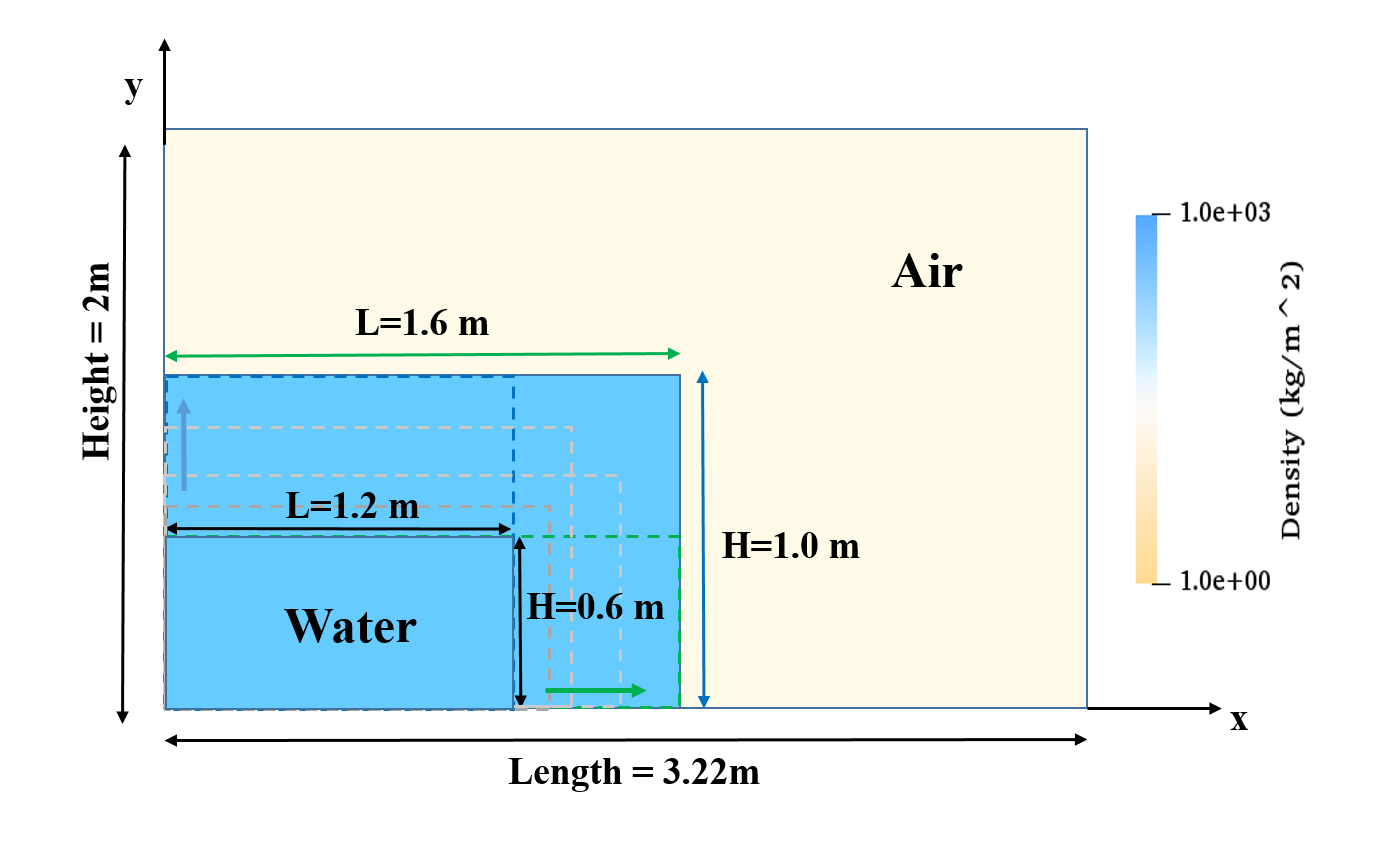}
\caption{The setttings of length and height in water collapse experiment.}
\label{inputs}
\end{figure}

{\it {Data collection}}: In this case, for training the hybrid DAA, the length $L$ and height $H$ of the tank are selected as the input parameters (as shown in \Cref{inputs}). The length of the tank $L$ is ranged from 1.2 $m$ to 1.6 $m$, and the height $H$ of the tank varies from 0.6 $m$ to 1.0 $m$. A series of sizes as inputs $\mu$ and the corresponding solution snapshots ${\hbar_d}$ (a total of 3375 snapshots of 45 pairs of input-output) were obtained by running the high fidelity model, which can be re-written in a time discretized form:
\begin{equation}
\mu = [{\mu ^1_a},{\mu ^2_a},...,{\mu ^{{N_x}}_a}], {(a \in [0, 44], {N_x}=19097)},
\end{equation}
\begin{equation}
\hbar_d = \left[ {\begin{array}{*{20}{c}}
{\hbar _{d,a,{t_1}}^1}&{\hbar _{d,a,{t_1}}^2}& \cdots &{\hbar _{d,a,{t_1}}^{{N_x}}}\\
{\hbar _{d,a,{t_2}}^1}&{\hbar _{d,a,{t_2}}^2}& \cdots &{\hbar _{d,a,{t_2}}^{{N_x}}}\\
\vdots & \vdots &{}& \vdots \\
{\hbar _{{d,a, t_{{N_t}}}}^1}&{\hbar _{{d,a,t_{{N_t},}}}^2}& \cdots &{\hbar _{{d,a,t_{{N_t},}}}^{{N_x}}}
\end{array}} \right], {(a \in [0, 44], {N_x}=19097, {N_t}=75)}.
\end{equation}

The training and validated input-output pairs ${\vartheta _{tr}}$ and ${\phi _{tr}}$ (where ${\mu_a \in {\mathbb{R}^{^{N_x}}}}, {\hbar_{d, a}\in {\mathbb{R}^{^{N_x \times N_t}}}}$, $a \in [0, 38], {N_x}=75, {N_t}=19097$) are selected for model training and parameter optimization, while the remained inputs ${(\vartheta \backslash {\vartheta _{tr}})}$ (where $ {\mu_a \in {\mathbb{R}^{^{N_x}}}, a \in [0, 6], {N_x}=19097}$) are used for model prediction.

\subsection{Model Prediction}
\label{sec5.1}

\begin{figure}[H]
\centering
\includegraphics[width=1.0\textwidth]{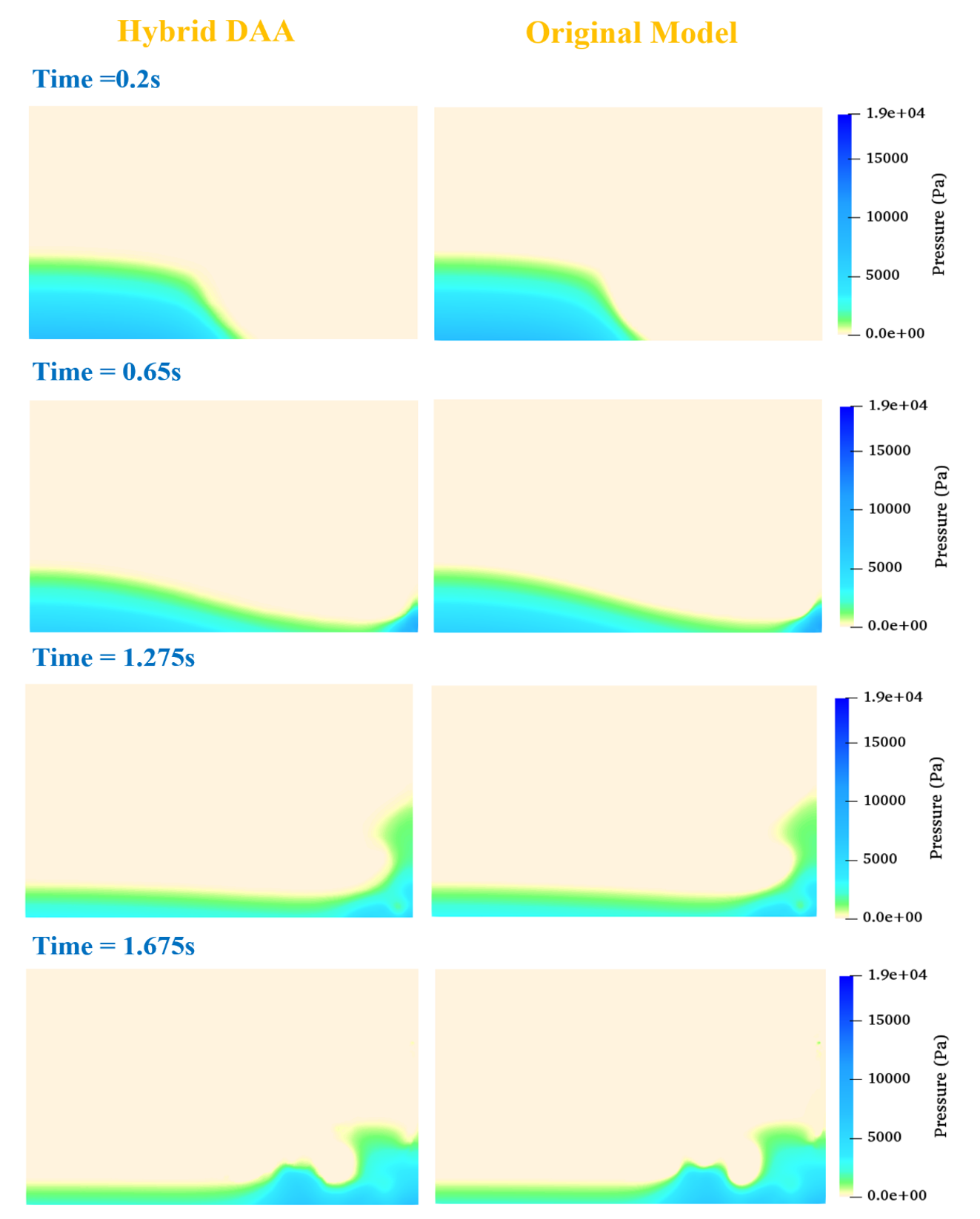}
\caption{Comparison of the spatial distribution of pressure fields obtained from the hybrid DAA (left) and the original high fidelity model (right) at time levels $t = 0.2, 0.65, 1.275, 1.675 s$.}
\label{fields}
\end{figure}

\begin{figure}[H]
\centering
\includegraphics[width=1.0\textwidth]{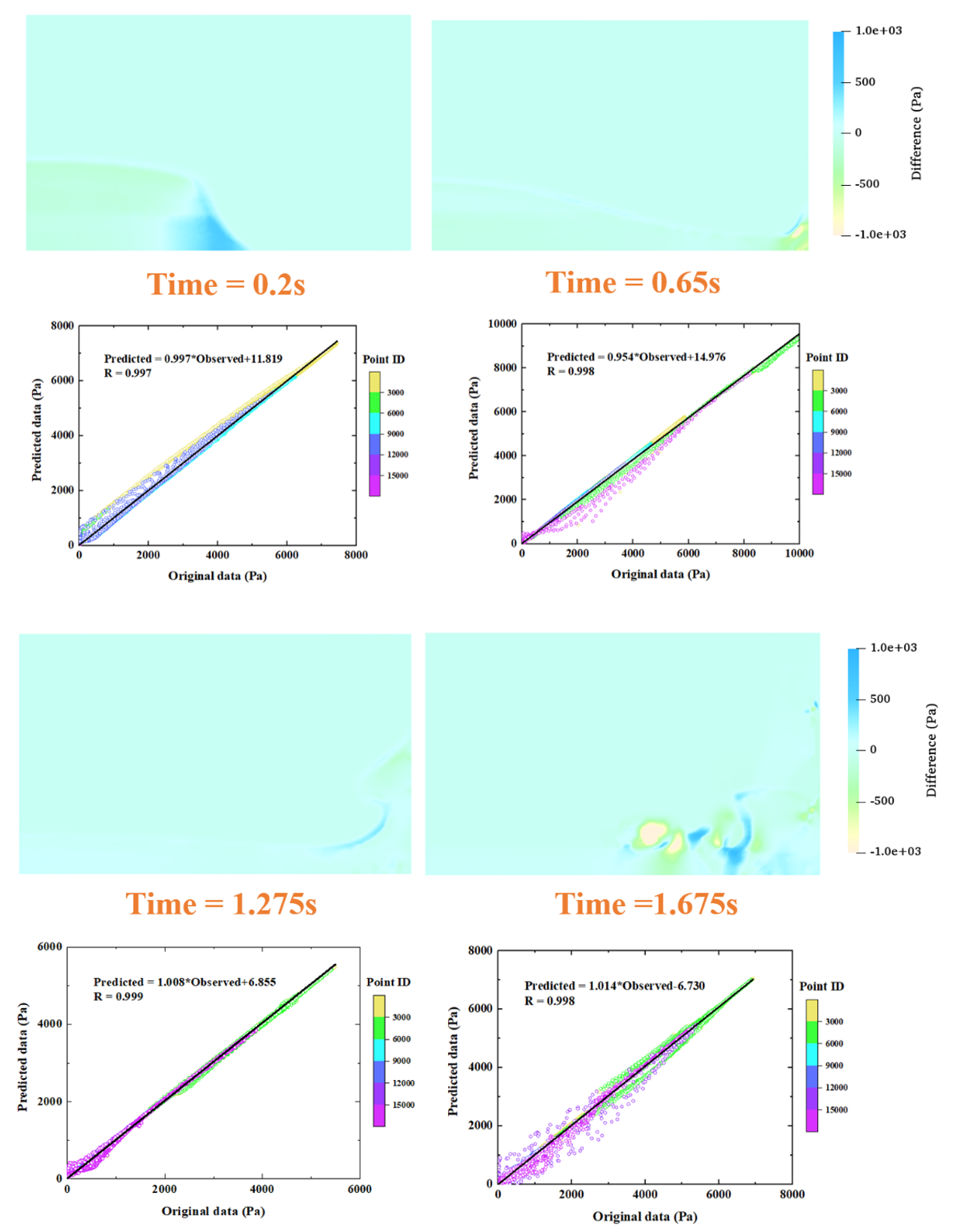}
\caption{Differences and correlation coefficients of pressure fields between the hybrid DAA and the original high fidelity model at time levels $t = 0.2, 0.65, 1.275, 1.675 s$.}
\label{error}
\end{figure}

{\it {Prediction of spatial nonliear fluid flows}}: To evaluate the predictive ability of the hybrid DAA, given the new input ${\mu \in (\vartheta \backslash {\vartheta _{tr}})}$, comparison of the predictive results from the hybrid DAA and the original high fidelity model are shown in \Cref{fields}. It can be seen that the hybrid DAA predicted the flow fields well, which captures the most pressure features at time levels $ t = 0.2, 0.65, 1.275, 1.675 s$. Visually, very little difference between the hybrid DAA and the original high fidelity model can be noticed. In order to compare the differences between the predictive and original fluid fields, the absolute error and correlation coefficient of pressure solutions within the computational domain area $\Omega$ are illustrated in \Cref{error}. It is observed that the absolute errors are small over the whole domain area at different time levels, and the correlation coefficient between the hybrid DAA and the high fidelity model is higher than $0.99$. The predictive results in spatial space suggest the hybrid DAA is able to obtain reasonable and accurate solutions in spatial space for nonlinear fluid flows.

\vfill
\clearpage

\begin{figure}[H]
\centering
\includegraphics[width=1.0\textwidth]{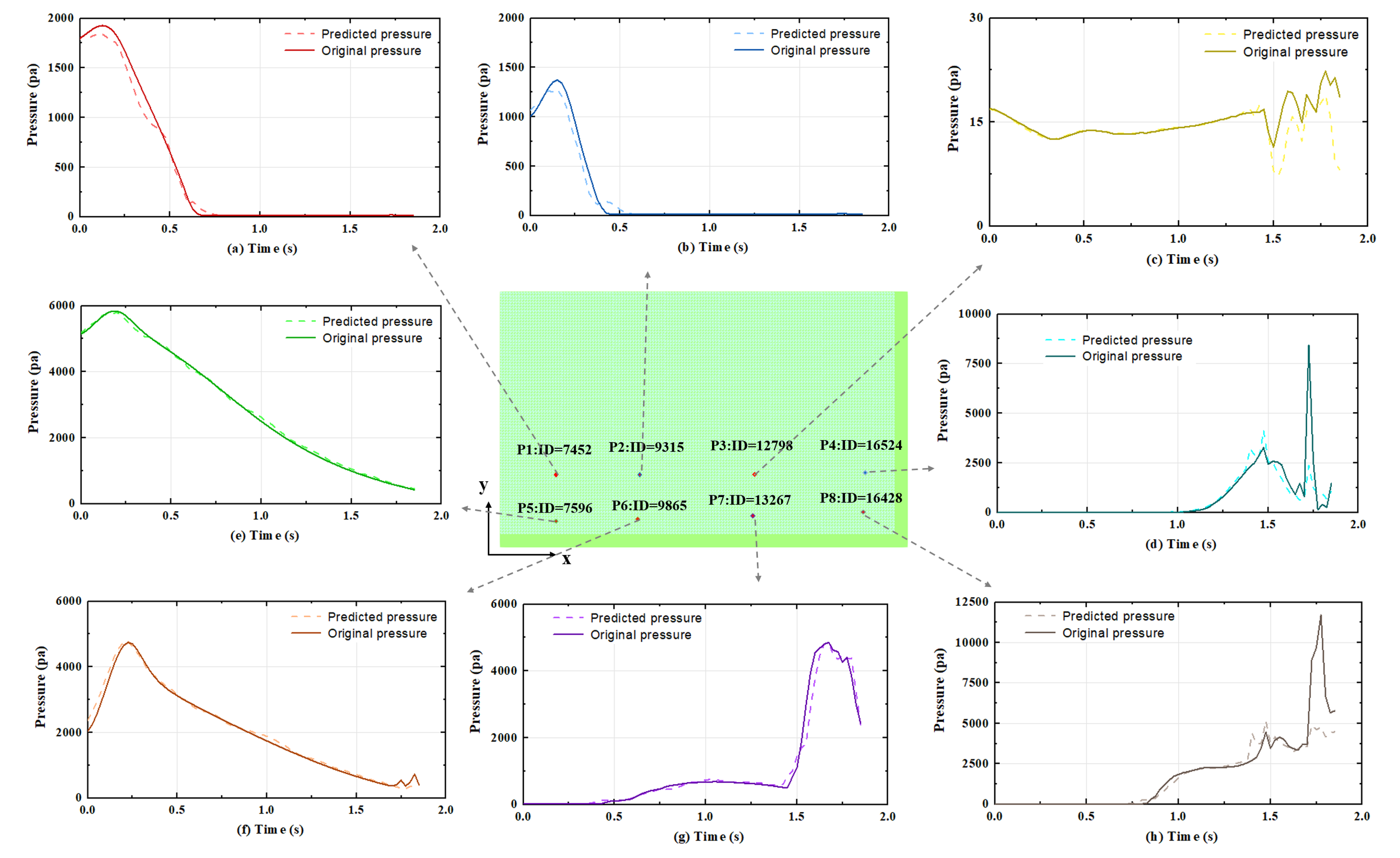}
\caption{Comparison of the temporal variation of pressure at points ID = 7452, 7596, 9315, 9865, 12798, 13267, 16428, 16524.}
\label{points}
\end{figure}

{\it {Prediction of temporal nonliear fluid flows}}: \Cref{points} shows the temporal variation of pressure solution predicted by the two models. Before the water collapse happens, the points P1, P2, P5 and P6 are located in the water area (as marked with a blue rectangle in \Cref{inputs}), while other four points P3, P4, P7 and P8 are placed in the area filled with air. Model performance has been evaluated by comparison of pressure variation in these detector points. In the \Cref{points}(a), (b), (e) and (f), the pressure values at four points are slightly increased and then gradually decreased as the water collapse happens. As for detector P3, it is beyond the scale of water front motions before $t = 1.5 s$, and the pressure trend in \Cref{points}(c) is not so obvious. After $t = 1.5 s$, influenced by the overturning water, the pressure at the detector P3 starts to fluctuate. Compared to detector P3, it can be noted that the pressure values at detector P4, P7 and P8 are uprising when the water drops down the horizon into rightward direction. The sequent water experiences dropping, traveling, rising up along the right vertical wall, and dropping again processes. It can be observed that the curves of pressure from the hybrid DAA achieve a good agreement with that of the original high fidelity model, except for the under-predictions of dropping again process after $t = 1.5 s$. However, it is also a challenge problem for numerical simulation when the water hits at the right wall and again meets the horizontal free surface \citep {park2009volume}. 

\vfill
\clearpage

\begin{figure}[H]
\centering
\includegraphics[width=0.8\textwidth]{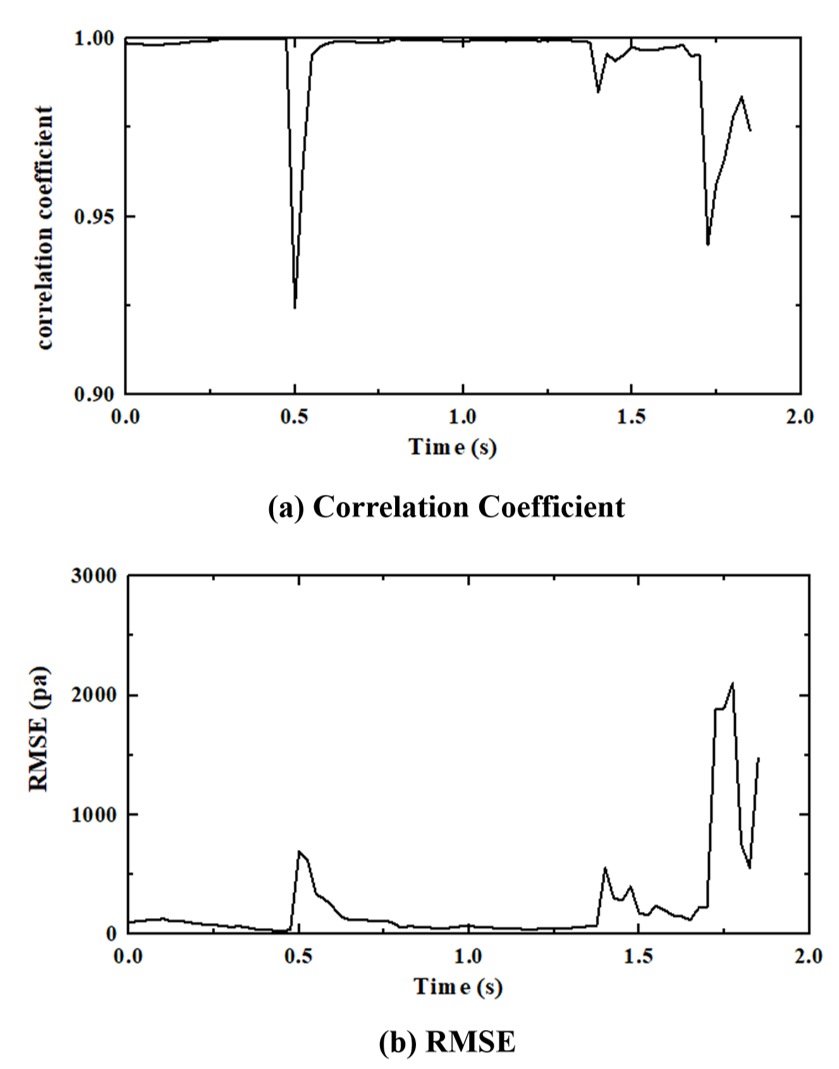}
\caption{The correlation coefficients and RMSE of pressure between the generated fields and original high fidelity fields during whole simulational period.}
\label{R}
\end{figure}

{\it {Model performance anaylsis}}: To further evaluate the performance of the hybrid DAA, the correlation coefficient and RMSE are calculated in temporal and spatial space, respectively. The error analysis of pressure solutions in temporal space is shown in \Cref{R}. It is evident that the values of RMSE are between $0 pa$ to $2000 pa$ while the correlation coefficients are above ${90\%}$. In light of fluid field prediction in temporal space, these results demonstrate that the hybrid DAA performs well and the predicted fluid fields are in good agreement with the true fluid fields.

\begin{figure}[H]
\centering
\includegraphics[width=0.8\textwidth]{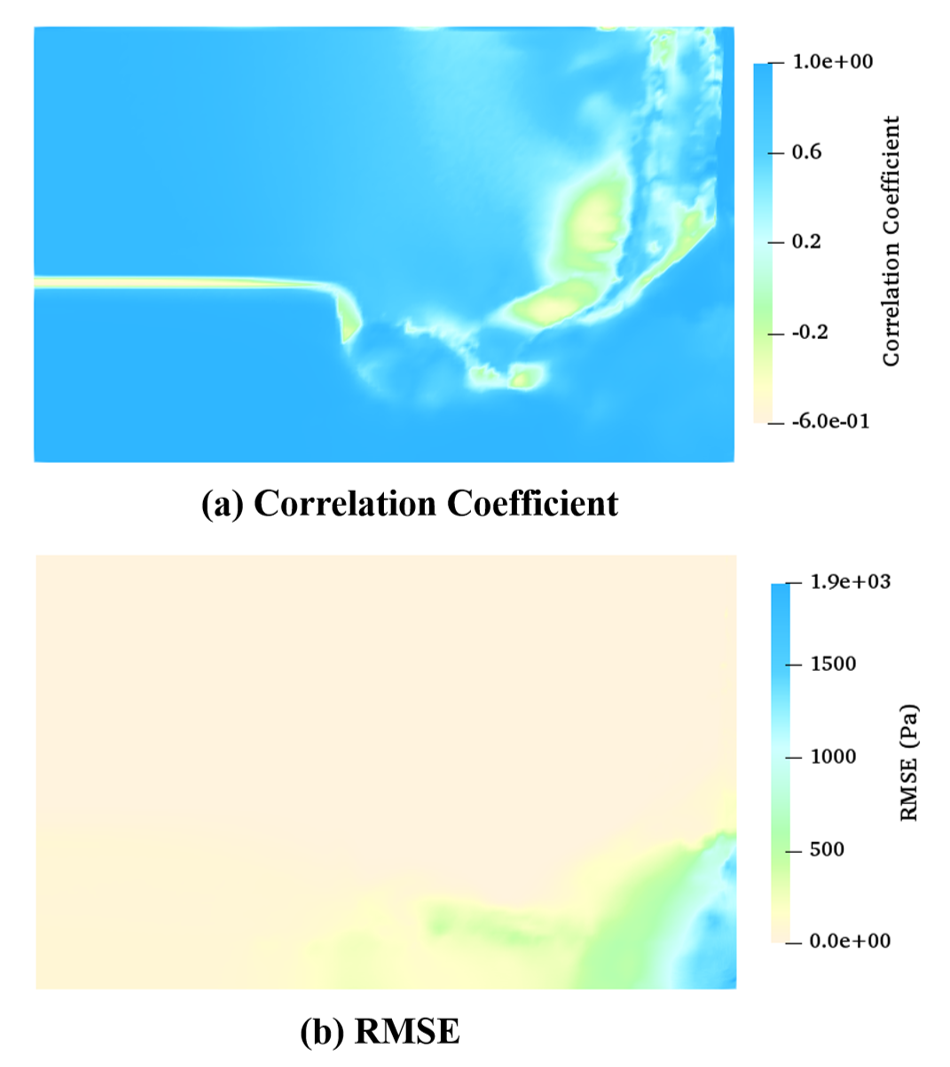}
\caption{The correlation coefficients and RMSE of pressure in time series in the computational domain area $\Omega$.}
\label{R2}
\end{figure}

\Cref{R2} illustrates the model performance in spatial space. It can be noted that the RMSE is not beyond $1900 pa$ in the whole domain area $\Omega$, which demonstrates the capability of hybrid DAA to predict accurate fluid flows. However, the correlation coefficient is negative in some area, where is affected by the water jet. When the water front strike the right vertical wall, the water jet causes a sudden rise of pressure and air entrainment \citep {park2009volume}. The phenomena is decribed in \Cref{points}(c), where depicts the pressure fluctuation after $t = 1.5 s$. Except for the area influenced by the interface of multi-materials, the correlation coefficient is beyond $0.9$ in the domain area $\Omega$.

%\vfill
%\clearpage

\subsection{Model efficiency}
\label{sec5.3}
The simulations of the hybrid DAA and the original high fidelity model were preformed on Intel Xeon(R) CPU@ 3.60GHz with a 449.5 GB memory. The computation time required for the hybrid DAA in online prediction process is $21.76 s$, while $8836.67 s$ for running the original high fidelity model. It can be seen that the CPU time for running the hybrid DAA is reduced drastically by three orders of magnitude in comparison to the original high fidelity model.

\section{Conclusions}
\label{sec6}
In this work, a hybrid DAA method has been, for the first time, used to predict nonlinear fluid flows in varying parameterized space. For any given input parameters $\mu$, the hybrid DAA is capable of predicting accurate dynamic nonlinear fluid flows and remains high efficiency in simulation, which takes both advantages of VAE and GAN. 

The performance of the hybrid DAA has been demonstrated by a water collapse test case. To evaluate the model accuracy, a detailed comparison between the original high fidelity model (Fluidity) and the hybrid DAA has been undertaken. The accuracy assessment has also been performed through the correlation coefficient and RMSE. The numerical simulations show that the hybrid DAA exhibits good agreement with the original high fidelity model, in both space and time. Additionally, a significant CPU speed-up has been achieved by the hybrid DAA. 

The hybrid DAA is an efficient and robust tool for parameterized modelling and prediction of nonlinear fluid flows. It provides a wide range of applications, for instance, risk response management in emergencies and natural hazards (e.g. flooding, dam break, etc), real-time decision making. Future work will be focused on the predictive ability of lead-time by the hybrid DAA and more complex fluid problems can also be examined with this model.

\section*{Acknowledgments}
The Authors acknowledge the support of: China Scholarship Council (No. 201806270238) and funding from the EPSRC (MAGIC) (EP/N010221/1) in the UK.

\end{spacing}
\renewcommand\refname{References}
\bibliographystyle{model4-names}
\bibliography{paper2}

\end{document}